
\input harvmac
%
%
%
%
%
\ifx\answ\bigans
\else
\output={
  \almostshipout{\leftline{\vbox{\pagebody\makefootline}}}\advancepag
eno
}
\fi
%
%
%
\def\mayer{\vbox{\sl\centerline{Department of Physics 0319}%
\centerline{University of California, San Diego}
\centerline{9500 Gilman Drive}
\centerline{La Jolla, CA 92093-0319}}}
%
%

%
%
\def\UCSD#1#2{\noindent#1\hfill #2%
\bigskip\supereject\global\hsize=\hsbody%
\footline={\hss\tenrm\folio\hss}}
%
%
\def\abstract#1{\centerline{\bf Abstract}\nobreak\medskip\nobreak\par
#1}
%
%
%
%
\edef\tfontsize{ scaled\magstep3}
 \tfontsize  \tfontsize
\font\titlermss=cmr5 \tfontsize \font\titlei=cmmi10 \tfontsize
\font\titleis=cmmi7 \tfontsize \font\titleiss=cmmi5 \tfontsize
\font\titlesy=cmsy10 \tfontsize \font\titlesys=cmsy7 \tfontsize
\font\titlesyss=cmsy5 \tfontsize  \tfontsize
\skewchar\titlei='177 \skewchar\titleis='177 \skewchar\titleiss='177
\skewchar\titlesy='60 \skewchar\titlesys='60 \skewchar\titlesyss='60
\scriptscriptfont0=\titlermss
\scriptscriptfont1=\titleiss
\scriptscriptfont2=\titlesyss
%
%
%
-
\def\inv{^{\raise.15ex\hbox{${\scriptscriptstyle -}$}\kern-.05em 1}}
\def\lbar{{\lower.35ex\hbox{$\mathchar'26$}\mkern-10mu\lambda}}

%
%
%
%
\def\dsl{\,\raise.15ex\hbox{/}\mkern-13.5mu D} 
subscripted
\def\delsl{\raise.15ex\hbox{/}\kern-.57em\partial}
\def\Ksl{\hbox{/\kern-.6000em\rm K}}
\def\Asl{\hbox{/\kern-.6500em \rm A}}
\def\Dsl{\hbox{/\kern-.6000em\rm D}} 
\def\Qsl{\hbox{/\kern-.6000em\rm Q}}
\def\gradsl{\hbox{/\kern-.6500em$\nabla$}}
%
%
\def\lspace{\ifx\answ\bigans{}\else\qquad\fi}
\def\lbspace{\ifx\answ\bigans{}\else\hskip-.2in\fi} 
%
%
\def\boxeqn#1{\vcenter{\vbox{\hrule\hbox{\vrule\kern3pt\vbox{\kern3pt
        \hbox{${\displaystyle #1}$}\kern3pt}\kern3pt\vrule}\hrule}}}
%
%
\def\mbox#1#2{\vcenter{\hrule \hbox{\vrule height#2in
\kern#1in \vrule} \hrule}}
%
%
%
%
\def\CA{{\cal A}}   \def\CD{{\cal
D}}

  \def\CO{{\cal O}} 
   
 \def\CV{{\cal V}}  
 
%
%
%
%
%

%

%
%

\def\darr#1{\raise1.5ex\hbox{$\leftrightarrow$}\mkern-16.5mu #1}

%
%
eqn
\def\frac#1#2{{\textstyle{#1\over #2}}} 
%
%
%
%

\def\Tr{\mathop{\rm Tr}}

%
%
%
%

%
%
\def\ltap{\ \raise.3ex\hbox{$<$\kern-.75em\lower1ex\hbox{$\sim$}}\ }
\def\gtap{\ \raise.3ex\hbox{$>$\kern-.75em\lower1ex\hbox{$\sim$}}\ }
\def\gl{\ \raise.5ex\hbox{$>$}\kern-.8em\lower.5ex\hbox{$<$}\ }
\def\roughly#1{\raise.3ex\hbox{$#1$\kern-.75em\lower1ex\hbox{$\sim$}}
}
%
%

%

%
\def\np#1#2#3{{Nucl. Phys. } B{#1} (#2) #3}

\def\prl#1#2#3{{Phys. Rev. Lett. } {#1} (#2) #3}

\relax

\def\CO{{\cal O}}

\def\lta{\ \hbox{\raise.55ex\hbox{$<$}} \!\!\!\!\!
\hbox{\raise-.5ex\hbox{$\sim$}}\ }
\def\gta{\ \hbox{\raise.55ex\hbox{$>$}} \!\!\!\!\!
\hbox{\raise-.5ex\hbox{$\sim$}}\ }
\def\mayer{\vbox{\sl\centerline{Department of Physics}
\centerline{University of California, San Diego}
\centerline{9500 Gilman Drive 0319}
\centerline{La Jolla, CA 92093-0319}}}

\def\ibid{{\it ibid.}}
\def\np#1#2#3{Nucl. Phys. {\bf B#1}, #3 (#2)}

\def\prl#1#2#3{Phys. Rev. Lett. {\bf #1}, #3 (#2)}
\def\bzc{\vec\xi^{\ (k)}_\alpha}
\def\bino#1#2{\left(\matrix{#1\cr#2\cr}\right)}
\def\txtbino#1#2{\textstyle{\left(\matrix{#1\cr#2\cr}\right)}}
\def\CAsl{\, \hbox{/\kern-.7000em $\CA$}\, }
\def\CVsl{\hbox{/\kern-.6500em $\CV$}}
\def\CDsl{\hbox{{\ {/\kern-.6500em $\CD$}}}}
\def\frac#1#2{{\textstyle{#1 \over #2}}}

\def\rmdp{{{\rm d}^{2n+1} p\over (2\pi)^{2n+1}}}

\def\epsi{\epsilon_{\alpha_1\beta_1\cdots\alpha_n\beta_n\alpha_{n+1}}
}
\def\epsii{\epsilon_{\mu_1\cdots \mu_{2n+1}} }

\def\({\left(}\def\){\right)}
\def\mayer{\vbox{\sl\centerline{Department of Physics}
\centerline{9500 Gilman Drive 0319}
\centerline{University of California, San Diego}
\centerline{La Jolla, CA 92093-0319, USA}}}

\def\[{\left[}
\def\]{\right]}
\def\({\left(}
\def\){\right)}

\noblackbox
\vskip 1.2in
\centerline{{\titlefont{Chern-Simons Currents and}}}
\vskip .2in
\centerline{{\titlefont{ Chiral Fermions  on the Lattice}}}
\bigskip
\centerline{\sl Submitted to Physical Review Letters}
\vskip .5in
\centerline{ Maarten F. L. Golterman}
\bigskip
\centerline{\sl Department of Physics}
\centerline{\sl Washington University}
\centerline{\sl St. Louis, MO 63130-4899, USA}
\bigskip\medskip
\centerline{\it and}
\bigskip\medskip
\centerline{Karl Jansen and
David B. Kaplan\footnote{$^{\dagger}$}
{Sloan Fellow, NSF Presidential
Young Investigator, and DOE Outstanding Junior
Investigator.}
}
\bigskip
\mayer
\vfill

\noindent
\line{E-mail:\hfill}
\line{maarten@wuphys.wustl.edu, jansen@higgs.ucsd.edu,
dkaplan@ucsd.bitnet\hfill}
\eject
\abstract{
We compute the Chern-Simons current induced by Wilson fermions on a
$d=2n+1$ dimensional lattice, making use of a topological
interpretation
of the momentum space fermion propagator as a map from the torus to
the
sphere, $T^{d}\to S^{d}$.  These mappings are shown to fall in
different homotopy classes depending on the value of $m/r$, where $m$
is
the fermion mass and $r$ is the  Wilson coupling.  As a result, the
induced Chern-Simons term changes discontinuously at $d+1$ different
values for $m$, unlike in the continuum. This behavior is exactly
what is
required by the
peculiar spectrum found for a recently proposed model of chiral
lattice
fermions as zeromodes bound to a domain wall.
   }
\vfill\UCSD{\hbox{UCSD/PTH 92-28,\ \break Wash.~U.~HEP/92-62}}{Aug
1992}

\eject

Recently a method for simulating chiral fermions on the lattice was
proposed by one of us \ref\david{D.B. Kaplan, {\it A Method for
Simulating Chiral Fermions on the Lattice}, UCSD-PTH-92-16, to appear
in
Phys. Lett. B.}.  The idea is to implement an odd dimensional
$d=2n+1$
theory of Wilson fermions with a mass coupling of the fermions  to a
domain wall. The low energy effective theory consists of massless
chiral
fermions bound to the $d-1$ dimensional domain wall, without there
being
``doubler'' modes
\ref\ksm{K.G. Wilson, in ``New Phenomena in Subnuclear Physics'', ed.
A.
Zichichi (Plenum, New York, 1977) (Erice, 1975);
L.H. Karsten and J. Smit, \np{183}{1981}{103}.}\ from the Brillouin
zone  boundary.  The anomalous Ward identities for these chiral
fermions in the presence of background gauge fields can be directly
measured on a finite lattice; this has already been done numerically
for
$d=3$ and the results are in agreement with the continuum anomaly in
$1+1$ dimensions \ref\karl{K. Jansen, {\it Chiral Fermions and
Anomalies on a Finite Lattice}, UCSD-PTH-92-18, (hep-lat/9209002)
 to appear in Phys. Lett.
 B.}.  In this system, the anomalous divergence of the $d-1$
dimensional
zeromode flavor currents is due to charge flow in the  direction
normal
to the domain wall, even though there is a mass gap off the wall---an
effect  discussed by Callan and Harvey \ref\ch{C.G. Callan, Jr.,
and J.A. Harvey, \np{250}{1985}{427}.}\ for continuum fermions
coupled to
a domain wall.  They pointed out that the Chern-Simons action induced
by
integrating out the heavy fermion modes \ref\oddam{J. Goldstone and
F.
Wilczek, \prl{47}{1981}{986}; S. Deser, R. Jackiw and S. Templeton,
Ann.
Phys. {\bf 140}, 372  (1982); \ibid\ {\bf 185},  406E (1988);   N.
Redlich, \prl{52}{1984}{18}; A.J. Niemi and G. Semenoff,
\prl{51}{1984}{2077}.}, being proportional to $m/\vert m\vert$,  has
opposite signs on the two sides of the domain wall.  This  gives rise
to
a Chern-Simons current in the presence of background gauge fields
with a nonzero divergence at the domain wall.  Furthermore, the
divergence  exactly reproduces the even $(d-1)$ dimensional anomaly
for
the single chiral fermion zeromode that is bound to the domain wall.
  In this Letter, we perform the Callan--Harvey (CH)
analysis for the lattice theory in Euclidian space, where the
zeromode
spectrum is more complicated than in the continuum.

It is far from obvious  that the lattice theory should follow the CH
continuum analysis; after all, the coefficient of the Chern-Simons
action gets $\CO(1)$ contributions from arbitrarily heavy fermion
modes,
and the heavy spectrum on the lattice looks nothing like in the
continuum.  In fact, we know the induced Chern-Simons operator must
have
a coefficient  very different from the continuum result.  While ref.
\david\ analyzed the  spectrum of the theory for a Wilson coupling
$r=1$
and a domain wall height $0< m_0  < 2$  and found a single chiral
mode,
a recent paper by Jansen and Schmaltz \ref\martin{ K. Jansen and M.
Schmaltz, {\it Critical Momenta of Lattice Chiral Fermions},
UCSD-PTH-92-29, submitted to Physics Letters
B.} analyzes the same model for general parameters and shows that the
spectrum bound to the domain wall changes
discontinuously with varying $ m_0 /r$ \ref\footi{All dimensionful
parameters are given in lattice units.  By a domain wall of height $
m_0$ we mean a spatially dependent mass term $m(s)\to \pm m_0$ as
$s\to
\pm\infty$, where $s$ is   the coordinate transverse to the domain
wall.}.  They find that for  $2k<\vert
m_0/r\vert<2k+2$---where $k$ is an integer in the range $0\le k\le
(d-1)$---there are  $\txtbino{d-1}{k}$ chiral modes bound to the
domain
wall with chirality $(-1)^k\times {\rm sign(m_0)}$; there are no
chiral
fermions for $\vert m_0/r\vert>2d$.
This is quite different than the continuum theory, for which there is
a
single chiral mode for any $ m_0\ne 0$.  If the induced Chern-Simons
action on the lattice is to correctly
account for the anomalous divergences of the chiral fermion currents
on
the domain wall, then evidently its coefficient must also depend
discontinuously on $m_0/r$ in a very particular way.  We show in this
Letter that that does indeed
happen \ref\footii{The dependence of the induced Chern-Simons action
on
the Wilson coupling $r$ has been previously discussed for three
dimensions in the continuum limit (spatially constant $m\to 0$) in {
 H. So,  Prog. Theor. Phys. {\bf 73},  528 (1985); {\bf 74}, 585
(1985);
and for $\vert m \vert<1$ by
{A. Coste and M. L\"uscher, \np{323}{1989}{631}.}  Some of the
techniques used in this Letter are similar to those found in the
latter
work.}}.

The  Abelian Chern-Simons action
in $d=(2n+1)$ continuous Euclidian dimensions is given by
\eqn\csop{\Gamma^{(d)}_{CS}=
\epsilon_{\alpha_1\cdots\alpha_{2n+1}} \int {\rm d}^{2n+1}x\
A_{\alpha_1}\partial_{\alpha_2}A_{\alpha_3}\cdots\partial_{\alpha_{2n
}}A
_{\alpha_{2n+1}}\
.}
 When a massive fermion is integrated out of the theory it generates
a
contribution to the effective action of the form
$S_{eff}=c_n\Gamma_{CS}$.  The coefficient $c_n$  can
be computed by calculating the relevant portion of the graph in
\fig\chern{The Feynman diagram
in $2n+1$ dimensions contributing to the induced Chern-Simons action
for
Abelian gauge fields; $\sum_{i=1}^{n+1} q_i = 0$.  Graphs with
multiple
photon vertices peculiar to
the lattice do not contribute, having the wrong Lorentz structure.}.
This is true on the lattice as well in the weak field, long
wavelength
limit for the gauge fields.
Denoting the fermion propagator and photon vertex as $S(p)$ and
$i\Lambda_\mu(p,p')$ respectively,  the graph of \chern\ yields a
value
for  $c_n$ which may be expressed as
\eqn\cval{\eqalign{\openup 4pt
c_n=&
{i\epsi\over(n+1)(2n+1)!} \left({\partial\ \ \over\partial
(q_1)_{\beta_1}}\right)\cdots \left({\partial\ \ \over\partial
(q_n)_{\beta_n}}\right)\times \cr
& \int_{\rm BZ} \rmdp\ \Tr\left[
S(p)\Lambda_{\alpha_1}(p,p-q_1)S(p-q_1)\cdots
\Lambda_{\alpha_{n+1}}(p+q_{n+1},p)\right]\
\biggl\vert_{q_i=0}\ .\biggr.}}
The factor of $(n+1)$ in equ. \cval\ is due to the symmetry factor of
the graph \chern; the
factor of $i$ is the product of $i^{n+1}$ from the photon vertices
and
$(-i)^n$ from relating the derivatives in equ. \csop\ to powers of
momenta.  The  $p$-integration is over the Brillouin zone of a
hypercubic lattice with lattice spacing $a=1$.

The integral \cval\ looks very difficult to compute on the lattice
for
arbitrary $n$, as both $S(p)$ and $\Lambda_\mu(p,p')$ are in general
rather complicated functions.  It is made quite tractable, however,
by
exposing its topological properties.  Gauge invariance implies that
the
photon coupling satisfies  the Ward  identity \ref\fotiii{We expand
the
lattice gauge field as $U_\mu(x) = 1-iA_\mu(x)+\ldots$.}
\eqn\wt{\Lambda_\mu(p,p)=-i{\partial\over{\partial p_\mu}}S^{-1}(p),}
allowing $c_n$ in equ.\cval\ to be reexpressed as
\eqn\cvalii{
c_n =  {{(-i)^n\epsii}\over(n+1)(2n+1)!} \int  \rmdp\ \Tr\left\{
\left[S(p)\partial_{\mu_1}S(p)^{-1}\right]\cdots
\left[S(p)\partial_{\mu_{2n+1}}S(p)^{-1}\right]\right\}}
where the differentiation $\partial_i$ is with respect to $p_i$.
The fermion propagator may be written in the generic form
\eqn\fprop{\eqalign{
S^{-1}(p) &= a(p) + i\vec b(p)\cdot\vec\gamma\cr
&= N(p) \left(\cos\vert\vec\theta(p)\vert +
i\hat\theta(p)\cdot\vec\gamma \sin\vert\vec\theta(p)\vert\right)\cr
&\equiv N(p) V(p)}}
where $N(p) \equiv \sqrt{a^2 + \vec b\cdot\vec b}$,
 $\ \ \vec\theta(p)=\hat b\arctan(\vert\vec b\vert/ a)$,
and $V(p)$ is seen to be a $2^n\times 2^n$ unitary matrix.
Provided that $S^{-1}(p)$ doesn't vanish for any $p$,
equ. \cvalii\ is independent of $N(p)$, allowing $S^{-1}(p)$ and
$S(p)$ to be replaced everywhere by $V(p)$ and $V^{\dagger}(p)$
respectively.
This matrix $V(p)$ is seen to   describe a mapping from
momentum space---which on the hypercubic lattice is the torus
$T^d$---onto the
sphere $S^d$ defined by the vector $\vec\theta(p)$. The homotopy
classes of such maps are identified by integers, and so the integral
in
equ. \cvalii\  has a simple topological interpretation:
it is, up to a normalization constant, the winding number of the map
described by the fermion propagator.

We now proceed to compute this winding number for the
 Wilson fermion propagator.  Although we are ultimately interested in
the effective action for lattice fermions in the presence of a domain
wall, we can compute the effective action far from the mass defect on
either side by
treating the fermion mass as a constant $m$.  Thus we can use the
standard Wilson propagator
\eqn\wilson{S^{-1}(p) = \sum_{\mu=1}^d i\gamma_\mu\sin p_\mu
+ m+ r \sum_{\mu=1}^d\left[\cos p_\mu -1\right]\ .}
Continuous changes in the mass and Wilson coupling, $m$ and $r$,
cannot
change the value of the winding number except at points for which
$S^{-1}$ has a zero for some momentum $p$.  Such singular points of
the
mapping occur only at momenta corresponding to the corners of
the Brillouin zone, and then only
for $m/r= 0,2,\cdots,2d$.
The Chern-Simons coefficient $c_n$ as a function of
$m/r$ must therefore be piecewise constant, changing only at these
critical values. Furthermore, for fixed $r$, $V(p)\to \pm 1$ as $m\to
\pm\infty$, so we may deduce that
\eqn\asymp{c_n(m,r)=0\qquad\ \ \  {\rm for\ } m/r<0\ ,\ \ \ m/r>2d\
.}
To
compute $c_n$ for $0<m/r<2d$, we need only evaluate the
 derivative of the integral in equ. \cvalii\ with respect to $m$
across the critical values $m/r=0,2\ldots,2d$.  This task is
simplified
by the fact that $c_n(m+dm,r)$ is unchanged as one deforms $dm$ in a
$p$-dependent way so that $dm(p)$
vanishes for all $p$ except in the vicinity of
the Brillouin zone corners;
these points are denoted by $\vec p =
\bzc$, the $\alpha= 1,\ldots,\txtbino{d}{k}$ vectors with $k$
nonvanishing components equal to $\pi$.
Therefore we need only evaluate the integrals in infinitesimal
regions
near the Brillouin zone corners $\bzc$.  After some algebra this
yields \ref\footiv{We work in $d=(2n+1)$ dimensions, and
our gamma matrix conventions are
$\{\gamma_\mu,\gamma_\nu\}=2\delta_{\mu\nu}$; $\gamma_\mu =
\gamma_\mu^{\dagger}$; $(\gamma_1\cdots\gamma_{d}) = i^n$.}
\eqn\dcval{\eqalign{
 {{\rm d} c_n(m,r)  \over {\rm d} m} &= {i(-1)^{n}2^n \over (n+1)}
\sum_{k,\alpha}
(-1)^k {{\rm d}\ \over{\rm d}m}\int_{{\rm d}\Omega}
\rmdp\  { (m-2rk)\over\left[\vert {\vec p}\vert^2 +
(m-2rk)^2\right]^{n+1}}\cr
&= {i(-1)^n\over(n+1)(2\pi)^n n!}\sum_{k=0}^{d}(-1)^k \bino{d}{k}
\delta(m-2rk)\ ,}}
where the integration ${\rm d}\Omega$ is over the infinitesimal
region
${\vert\vec p\vert<\epsilon\to 0}$.
This may be trivially integrated with respect to $m$, given the
boundary
values \asymp, to
yield the Chern-Simons coefficient for Wilson fermions:
\eqn\ivalf{ c_n(m,r) =
{i(-1)^n\over2(n+1)(2\pi)^n n!}\sum_{k=0}^{d}(-1)^k
\bino{d}{k} {m-2rk\over\vert m-2rk\vert}\ .\qquad{\rm (Wilson\
fermions)}}
This is to be compared to the continuum result for a fermion of mass
$m$, computed by inserting the continuum free propagator in equ.
\cval\
and integrating over continuum momentum space:
\eqn\conti{c_n(m) = {i(-1)^n\over2(n+1) (2\pi)^n n!}{m\over\vert
m\vert}
\ .\qquad\qquad{\rm (continuum\ result)}}
Our computation is  exact in the limit of small and
adiabatic external gauge fields, even for $m$ of $\CO(1)$ in lattice
units; it is readily generalized to non-Abelian gauge fields.

So far we have only considered lattice fermions with a constant mass;
our primary interest though is in studying current flow in the
presence
of a domain wall  mass term for the fermion.  Following Callan and
Harvey, we assume that sufficiently far away from the domain wall on
either side, we can treat the mass as constant and compute the
induced
current by varying the Chern-Simons action,
\eqn\jval{J_\mu^{CS}={\delta S_{eff}\over\delta A_\mu} = {(n+1)\over
2^n}  c_n(m(s),r)
\epsilon_{\mu\alpha_1\alpha_2\cdots\alpha_{2n+1} }
F_{\alpha_1\alpha_2\cdots} F_{\alpha_{2n-1}\alpha_{2n}}\ ,}
where the factor of $2^n$ arises from replacing $\partial A$ by the
field strength $F$.  Since $c_n$ depends on the mass $m(s)$ which is
space dependent, ${J_\mu}^{CS}$ has a nonzero divergence
corresponding
to current flow normal to the domain wall. In their continuum
analysis,
Callan and Harvey pointed out  that   with $c_n$ given by equ.
\conti,
there was equal current flow toward the
 wall from each side that exactly compensated for the anomalous
divergence of the chiral fermion current along the wall's surface.
For
lattice fermions in the presence of the domain wall, we must use the
value \ivalf\ for $c_n(m,r)$. Thus on the side of the domain wall for
which $m(s)/r$ is negative, the Chern-Simons current vanishes. On the
other side, where $m(s)/r\to \vert m_0/r\vert$, the current either
vanishes if $\vert m_0/r\vert > 2(d+1)$,  or else, for  $ 2\ell <
\vert
m_0/r\vert <2(\ell+1) $, the current is given by
\eqn\ratio{\eqalign{
{J_\mu^{CS}({\rm lattice})\over J_\mu^{CS}(\rm continuum)}
&= \sum_{k=0}^{\ell}(-1)^k \bino{d}{k}
-\sum_{k=\ell+1}^{d}(-1)^k \bino{d}{k} \cr
  &=2(-1)^{\ell}\bino{d-1}{\ell}\ .}}
  For example, for $\ell = 0$ (a domain wall height satisfying
$0<\vert
m_0/r\vert < 2$) the lattice current vanishes on one side of the
wall,
while having twice the continuum magnitude on the other side; in this
case the total divergence of the continuum and lattice currents are
the
same. In general the lattice Chern-Simons current vanishes on the
side
of the domain wall for which $m(s)/r$ is negative, while on the other
side has the correct magnitude to compensate for the anomaly of not
{\it
one} positive chirality zeromode at the domain wall, but rather
$\txtbino{d-1}{\ell}$
zeromodes  with chirality $(-1)^{\ell}\times {\rm sign} (m_0/r)$.
This
result agrees precisely with the
zeromode spectrum bound to the lattice domain wall found by Jansen
and
Schmaltz \martin.  It also agrees with a numerical computation of the
Chern-Simons current on a $d=3$ finite lattice that we have performed
(see ref. \karl), which exhibits the behaviour peculiar to the
lattice
that the Chern-Simons current flows on only one side of the domain
wall,
as well as the discontinuous dependence of its magnitude on $m_0/r$.

Our analysis confirms that the model of ref. \david\ correctly
reproduces the continuum anomaly for chiral fermions on a finite
lattice
in the presence of weak gauge fields, even for a domain wall height
$m_0
= \CO(1)$ in lattice units.

\vfill\eject
\centerline{\bf Acknowledgements}
\vskip.5in
We would like to thank J. Kuti,  A. Manohar, G. Moore, J. Rabin and
M.
Schmaltz for useful conversations.  M.G. would like to thank the T8
Division of  LANL and the Physics Departments of UC San Diego and UC
Santa Barbara for hospitality.

\bigskip
\noindent
M.G. is supported in part by the Department of Energy under grant
\#DE-2FG02-91ER40628.
K.J. and D.K. are
supported in part by the Department of Energy under grant
\#DE-FGO3-90ER40546, and D.K.  by the NSF under contract PHY-9057135,
 and by a fellowship from the Alfred P. Sloan Foundation.

\listrefs
\listfigs

\bye